\documentclass{article}
\usepackage{arxiv}
\usepackage[utf8]{inputenc} 
\usepackage[T1]{fontenc}    
\usepackage{url}            
\usepackage{booktabs}       
\usepackage{amsfonts}       
\usepackage{nicefrac}       
\usepackage{microtype}      
\usepackage{graphicx}
\usepackage{amsmath}
\usepackage{subfig}
\usepackage{caption}
\usepackage{booktabs,siunitx}
\usepackage{wrapfig,lipsum}
\usepackage{multirow}
\usepackage[]{nohyperref} 
\usepackage{url} 
\usepackage[dvipsnames,tabularx]{xcolor}
\usepackage{algorithm}
\usepackage{algorithmic}
\usepackage{rotating}
\usepackage{tabularx}
\usepackage{rotating}
\usepackage{float}
\usepackage{placeins}

\newtheorem{theorem}{Theorem}[section]

\newtheorem{corollary}[theorem]{Corollary}
\newtheorem{definition}{Definition}[section]

\DeclareMathOperator*{\argmax}{arg\,max} 

\title{ABCD: Algorithm for Balanced Component Discovery in Signed Networks}

\author{Muhieddine Shebaro and  Jelena Te\v{s}i\'{c}}

\begin{document}


\maketitle

\begin{abstract}
The largest balanced element in signed graphs plays a vital role in helping researchers understand the fundamental structure of the graph, as it reveals valuable information about the complex relationships between vertices in the network. The challenge is an NP-hard problem; there is no current baseline to evaluate state-of-the-art signed graphs derived from real networks. In this paper, we propose a scalable state-of-the-art approach for the maximum balanced sub-graph detection in the network of \emph{any} size. The proposed approach finds the largest balanced sub-graph by considering only the top $K$ balanced states with the lowest frustration index. We show that the ABCD method selects a subset from an extensive signed network with millions of vertices and edges, and the size of the discovered subset is double that of the state-of-the-art in a similar time frame.
\end{abstract}
balanced sub-graph, frustration index, balanced states, and signed graphs

\section{Introduction} 
\label{sec-Intro}
Signed networks allow for unsigned and negative weights in the graph-based representation. It is a graph where each edge between nodes is assigned a positive or negative sign. Negative weights represent antagonistic relationships and model the conflicting opinions between the vertices well \cite{2022wusurvey}. Balance theory represents a theory of changes in attitudes \cite{1958Abelson}: people's attitudes evolve in networks so that friends of a friend will likely become friends, and so will enemies of an enemy \cite{1958Abelson}. Heider established the foundation for social balance theory \cite{Heider}, and Harary established the mathematical foundation for signed graphs and introduced the k-way balance \cite{Har2,Harary1968}. The solutions to the tasks to predict edge sentiment, to recommend content and products, or to identify unusual trends have had balanced theory at its core \cite{derr2020link,garimella2021political,interian2022network,amelkin2019fighting}. The task of the \textbf{most extensive balanced sub-graph} discovery has applications in portfolio system's economic risk management \cite{Harary2002}, computational and operational research \cite{Figueiredo2014}, community analysis and structure \cite{Macon2012}, computational biology to model balanced interactions between genes and proteins \cite{Liu2022} and social network analysis \cite{Chen2023}. The vertices that are part of the maximum balanced sub-graph $\Sigma'$ of $\Sigma$ may not necessarily have a high degree of centrality between them. By locating the largest balanced sub-graphs, we can simplify the system into sub-systems with balanced interactions and eliminate inconsistencies regarding unbalanced cycles. The largest balanced sub-graph is the largest possible subset of the signed network that satisfies the balance theory (every cycle has an even number of negative edges). This sub-graph doesn't have to be a \textbf{clique} where a clique is a subset of nodes in the graph where every node has an edge over every other node in that subset.

Finding the largest balanced sub-graph is a well-known NP-hard problem \cite{2012Zaslavsky}, and existing solutions do not scale to real-world graphs \cite{2022wusurvey}. This paper proposes an algorithm for balanced component discovery (ABCD) in signed graphs, and we show that it discovers \emph{larger} signed sub-graphs faster than TIMBAL. The approach builds on the scalable discovery of fundamental cycles in \cite{2021Alabandi} and utilizes the graph's vertex density distribution and stable states to minimize the number of vertices removed from the balanced sub-graph. Section~\ref{sec-Prelim} explains the notations, definitions, and theorems behind the signed graph balancing and the algorithm for the scalable graph cycle-basis computation of the underlying unsigned graph $G$ of $\Sigma$. Section~\ref{sec-Method} introduces the novel ABCD algorithm and the implementation details. We use the edge sign switching technique using a fundamental cycle basis discovery method to \emph{search} for the maximum balanced sub-graph. In Section~\ref{sec-complexity}, we analyze the complexity of our algorithm. Section~\ref{sec-Exp} compares the proposed method to the state-of-art method proposed in \cite{TIMBAL}. The TIMBAL method achieved the highest vertex cardinality (number of vertices in the largest balanced sub-graph) across all signed graphs, among other baselines in the literature \cite{TIMBAL}. We evaluate our algorithm on the Konect and Amazon signed graphs in Subsection~\ref{ssec-KonectResults} and Subsection~\ref{ssec-AmazonResults}, respectively. Next, we perform an ablation study on our proposed algorithm in Subsection~\ref{ssec-param}. In Section~\ref{sec-Done}, we summarize our findings.

\textbf{Problem Definition:} It is to find the largest balanced component $\Sigma^G, |\Sigma^G|= g$ in \emph{any} size signed graph $\Sigma, |\Sigma|=n$ is in Equation~\ref{eq-subgraphB}. 
\begin{equation}
\Sigma^G \subseteq \Sigma \land Fr(\Sigma^G)=0 \land \argmax_{g \leq n }{\Sigma^G} \implies \Sigma^G
\label{eq-subgraphB}
\end{equation}

\noindent where $Fr(\Sigma^G)$ is the frustration of balanced sub-graph $\Sigma^G$ defined in \cite{2021Cloud}. The frustration is the level of imbalance (number of fundamental cycles with an odd number of negative signs) found in the network. 

\textbf{Goal:} It is to find a sub-graph in a signed network with an even number of negative edges along each fundamental cycle, and its size (node cardinality) is as large as possible.

 \begin{figure*}[!ht]
 \centering
 \includegraphics[width=\linewidth]{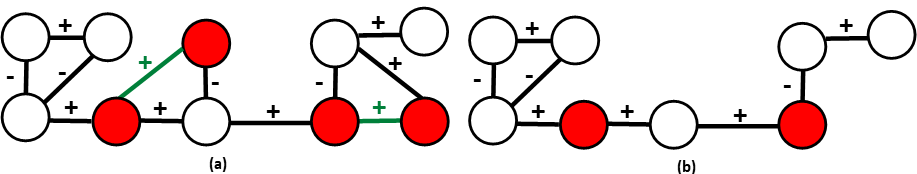}
 \caption{(a): The unbalanced signed network. Green edges are the candidate edges causing imbalance, and red vertices are the candidate vertices. (b): The maximum balanced signed sub-graph obtained after deleting one candidate vertex along each edge.}
 \label{fig-example_MBS}
\end{figure*}

\section{Definitions \& Corollaries}
\label{sec-Prelim}
In this section, we define the fundamental cycle basis and relevant signed graph network terms and outline the theorems and corollaries for the proposed ABCD approach. Table~\ref{tab-Notations} outlines the meaning of the notations used when describing the algorithm steps in the paper.
\begin{definition}
\textbf{Signed graph} $\Sigma=(G, \sigma)$ consists of underlying unsigned graph $G$ and an edge signing function $\sigma : e \rightarrow \{+1,-1\}$ \cite{signedsocial}. The edge $|E|$ can be positive $e^+$ or negative $e^-$. \textbf{Sign} of a sub-graph is \emph{product} of the edges signs. \textbf{Path} is a sequence of distinct edges $|E|$ that connect a sequence of distinct vertices $|V|$ in a graph. \textbf{Connected graph} has a path that joins any two vertices. \textbf{Cycle} is a path that begins and ends at the same node. \label{def-SignedGraph} 
\end{definition}

\begin{table}[!ht]
\caption{Summary of notations and their meanings.}
\label{tab-Notations}
\centering
\begin{tabular}{p{2cm}p{5.5cm}} \toprule
\textbf{Notation} & \textbf{Meaning} \\
\midrule
$\Sigma$/ G & signed network/ its underlying graph \\
$\Sigma^G$ & sub-graph \\
$T$/ $I$ & spanning tree/ \# of spanning trees\\
$v$/ $V_x$/ $V$ / $|V|$ & a node/ set of some vertices/ set of all vertices/number of nodes\\
$E_x$/$E$/ $|E|$ & a set of edges/ set of all edges in $\Sigma$ / number of edges \\
$e$/ $e^+$ / $e^-$ & an edge/ positive edge/ negative edge\\
$\Sigma'$ & \emph{any} balanced state of $\Sigma$\\
$\Sigma_i$ & $i^{th}$ nearest balanced state $\Sigma$\\
$Fr(\Sigma_i)$ & frustration from $\Sigma$ to $\Sigma_i$\\
$(U,W)$ & Harary bipartition sets, $|U| \geq |W|$\\
$\mathcal{F}_{\Sigma}$ & the frustration cloud set of $\Sigma$\\
$|\mathcal{F}_{\Sigma}|$ & the size of the frustration cloud.\\
$K$ & \# of nearest balanced states w lowest frustration index, $K \leq |\mathcal{F}_{\Sigma}|$\\
$H$ & binary array of size$|V|$ to separate the vertices into Harary bipartitions where $H[v] = 1$ if $v$ is in set $U$, and 0 if $v$ is in set $W$\\
$\mathcal{H}_{\Sigma}$ & container to save the collection of $H$ array for the top-$K$ balanced states with the lowest frustration\\
$\mathcal{E}_{\Sigma}$ & container of $\Sigma$ for storing a set of edges in each element that switched signs during balancing \\
$ \mathcal{F}_{\Sigma}$ & $K$ size array of the number of edge sign switches of top $K$ nearest balanced states.\\
$\mathcal{V}_i$& set of vertices to remove from graph $i$\\
$\mathcal{ABCD}$& a set containing $K$ sub-graphs after $V/\mathcal{V}_i$\\
\bottomrule
\end{tabular}
\end{table}

\begin{definition}
Graph $\Sigma^G$ is a \textbf{sub-graph} of a graph $\Sigma$ if \textbf{all} edges and vertices of $\Sigma^G$ are contained in $\Sigma$. \label{def-Subgraph} 
\end{definition}

\begin{definition}
\textbf{Balanced Signed graph} is a signed graph where every cycle is positive. The \textbf{Frustration Index} of a signed graph is the minimum number of candidate edges whose sign needs to be switched for the graph to reach the balanced state \cite{frustcite}. 
\end{definition}

\begin{definition}
\textbf{Cycle Basis} is a set of simple cycles that forms a basis of the cycle space \cite{Berger2004}. A \textbf{Bridge node} is a vertex whose removal increases the number of connected components within the network. \label{def-CycleBasis} 
\end{definition}

\begin{definition}
For the underlying graph $G$, let $T$ be the spanning tree of $G$, and let an edge $e$ be an edge in $G$ between vertices $x$ and $y$ that is \emph{NOT} in the spanning tree $T$. Since the spanning tree spans all vertices, a unique path in $T$ between vertices $x$ and $y$ does not include $e$. \textbf{The fundamental cycle} is any cycle that is built using path in $T$ plus edge $e$ in graph $G$. \label{def-FundamentalCycle} 
\end{definition}

\begin{corollary}
All the cycles formed by combining a path in the tree and a single edge outside the tree create a fundamental cycle basis from a spanning tree. Thus, the underlying unsigned graph $G$ with $|V|$ vertices and $|E|$ edges has precisely $|V|-|E|+1$ fundamental cycles.
\label{col-fundCycles}
\end{corollary}

\begin{definition}
The balanced states are \textbf{near-balanced} if and only if the original graph requires a minimum number of edge sign switches to reach a balanced state. We label the stable states of $\Sigma$ as $\Sigma_i$, where $i \in [1, |\mathcal{F}_\Sigma|]$. $|\mathcal{F}_\Sigma|$ is the the size of the frustration cloud in \cite{2021Cloud}\label{def-near} 
\end{definition}

\begin{theorem}If a signed sub-graph $\Sigma^G$ is balanced, the following are equivalent \cite{Har2}:
\begin{enumerate}
 \setlength{\leftmargin}{0pt}
 \item $\Sigma^G$ is balanced. (All circles are positive.)
 \item For every vertex pair $(v_i,v_j)$ in $\Sigma'$, all $(v_i,v_j)$-paths have the same sign.
 \item $Fr(\Sigma^G) = 0$.
 \item There exists a bipartition of the vertex set into sets $U$ and $W$ such that an edge is negative if, and only if, it has one vertex in $U$ and one in $W$. The bipartition ($U$,$W$) is called the \emph{Harary-bipartition}. Note the sets so that $U$ always contains a more significant or equal number of vertices than $W$. 
\end{enumerate}
\label{the-HararyCut}
\end{theorem}

\section{Related Work}
\label{sec-Related}

Finding the stable sub-graph in a signed graph is known to be an NP-hard problem. G\"{u}lpinar et al. proposed the GGMZ algorithm. GGMZ computes the input graph's minimum spanning tree, then selects the subset of vertices so that all the edges crossing that subset are inverted to create positive edges, and the result is the set of vertices disconnected by negative edges. The system's overall complexity is $O(|V|^3)$ if $V$ is a set of vertices \cite{GULPINAR2004359}. Poljak and Daniel Turz\'{i}k show that any signed graph that has $|V|$ vertices and $|E|$ edges contains a balanced sub-graph with at least $0.5|E| + 0.25(|V|- 1)$ edges \cite{POLJAK1986}. Crowston et al. \cite{Crowstown2013} propose a discovery of a balanced sub-graph of size $0.5|E| + 0.25(|V|- 1+k$) where $k$ is the parameter. They reduced data by finding small separators and a novel gadget construction scheme. Figueiredo et al. introduced a polyhedral-based branch-and-cut algorithm to find the largest sub-graph \cite{Figueiredo2011}. Then, they proposed GRASP, an improved algorithm version with the pre-processing and heuristic methods \cite{Figueiredo2014}. The GRASP algorithm randomly selects a subset of vertices. Greedily adds vertices that maximize the number of edges connecting them to the current subset while keeping the size of the subset balanced \cite{Figueiredo2014}. The EIGEN algorithm \cite{bonchi2019discovering} works by first computing the eigenvectors of the Laplacian matrix of the graph. Using the dominant eigenvector of the adjacency matrix, it then partitions the graph into two disjoint sets. The partition is made by setting a threshold value for the eigenvector and assigning each vertex to one of the two sets based on whether its value in the eigenvector is above or below the threshold. The algorithm then recursively this partitioning process on each of the two sets. Sharma et al. proposed a heuristic that deletes edges from the graph associated with the smallest eigenvalues in the Laplacian matrix of the graph until a maximum balanced sub-graph is obtained \cite{2021Sharma}. Ordozgoiti et al. introduced the most scalable version of the algorithm to date. TIMBAL is an acronym for \emph{trimming iteratively to maximize balance} two-stage method approach where the first stage removes vertices and the second one restores them as long as it does not cause imbalance \cite{TIMBAL}. Both algorithms rely on signed spectral theory. The approaches do not scale to the large signed graphs as they rely on the costly eigenvalue computation ($O(|E|^2)$), and its performance decreases due to the spectral pollution in eigenvalue computation \cite{BOULTON20161}. TIMBAL proposes a novel bound for perturbations of the graph Laplacian pre-processing techniques to scale the processing for large graphs. They evaluate the scalability of the proposed work on graphs over 30 million edges by artificially implanting balanced sub-graphs of a specific size and recovering them \cite{TIMBAL}. 

Note that the Maximizing balance via edge deletions (MBED) task is \emph{different} than the task of discovering the maximal balanced sub-graph. MBED task requires the target community and the budget as input, and the goal is to remove edges to make that input community as close to being balanced as possible \cite{2021Sharma}. Discovering the maximal balanced sub-graph \emph{does not require} community and budget specifications. 

Preserving connectivity when deleting vertices is critical to maximizing the chances of obtaining a large, balanced sub-graph. Kleinberg et al. \cite{conn1} considered a model for tracking the network connectivity under vertex or edge deletions, focusing on detecting $(\epsilon,k)$-failures. These failures occur when an adversary deletes up to $k$ network elements, each at least an $\epsilon$ fraction of the network, becoming disconnected. A set of vertices is called an $(\epsilon,k)$-detection set if, for any $\frac{1}{\epsilon}$-failure, some two vertices in that set of vertices can no longer communicate. The authors show that for an adversary that can delete $k\lambda$ edges, the random sampling approach can detect a set of size O($\frac{k}{\epsilon}log\frac{1}{\epsilon}$), and polynomial time is required to detect the $(\epsilon,k)$ set of minimum size with the proposed approach \cite{conn1}. 

Deng et al. \cite{Deng_2007} studied the effect of vertex deletion on the network structure by proposing an evolving network model. They revealed that as the intensity of vertex deletions increases, the network's degree distribution shifts from a scale-free to an exponential form and that vertex deletions generally decrease the network's clustering coefficient. On the other hand, the problem opposite to preserving connectivity upon vertex deletion is called the Critical vertex Detection Problem \cite{LALOU201892}. This problem has garnered significant interest recently, and the goal is to identify a set of vertices whose removal most effectively disrupts network connectivity based on specific connectivity metrics.

\FloatBarrier
\begin{figure*}[!h]
 \centering
 \rotatebox{90}{
   \begin{minipage}{\textheight}
     \centering
     \includegraphics[width=\linewidth]{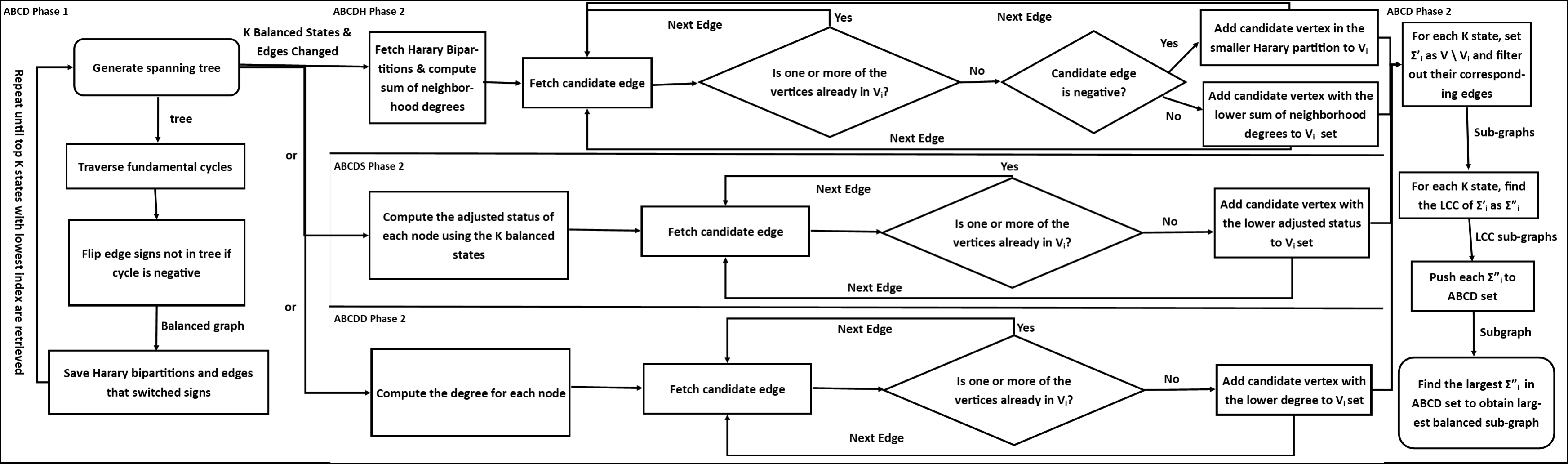}
     \caption{The ABCD pipeline.} \label{fig-pipeline}
   \end{minipage}
 }
\end{figure*}
\FloatBarrier

\section{Methodology}
\label{sec-Method} 

The \textbf{Algorithm for the Balanced Component Discovery (ABCD)} approach removes the minimal number of vertices by removing one vertex per imbalanced edge in a signed graph. We find a candidate set of imbalanced edges for the given signed graph by implementing a fast balancing algorithm proposed in \cite{2021Alabandi,2021Cloud}. Section~\ref{ssec-ABCD1} outlines the ABCD Phase 1 approach of collecting multiple candidate edge sets for deletion. Section~\ref{ssec-ABCD2} describes three different Algorithms for the Balanced Component Discovery (ABCD) approaches to approximate connectivity in the vertex removal procedure: Degree-based (ABCDD), Harary-based (ABCDH), and Status-based (ABCDS). We title Phase 1 as ``Top $K$ Nearest Balanced States Extraction and Candidate Edge Identification'' and Phase 2 as ``Candidate Vertex Purging and Largest Sub-graph Retrieval.'' Figure~\ref{fig-example_MBS} demonstrates an example execution of our algorithm where the balancing algorithm identifies the candidate edges causing imbalance (green) \cite{2021Cloud}. One of the red vertices along these candidate edges has been removed based on handling criteria.

\subsection{ABCD Phase 1} 
\label{ssec-ABCD1}

\begin{algorithm}[!ht]
\caption{ABCD Phase 1}
\label{alg-ABCD1} 
\begin{algorithmic}[1] 
\STATE Fetch signed graph $\Sigma$, number of iterations $I$, and integer $K$ that determines the stable states with the lowest frustration index to keep
\STATE Generate set of $I$ spanning trees $T$ of $\Sigma$
\STATE Create empty sets $ \mathcal{F}_{\Sigma}$, $\mathcal{E}_{\Sigma}$, and $\mathcal{H}_{\Sigma}$
\STATE Initialize $c$ = 0
\FOR{$i$ = 0; $i$++; $i$ < I}
\STATE Create empty set $M_i$
 \FOR{edges $e$, $e \in \Sigma \setminus T_i$}
 \IF{fundamental cycle $T_i \cup e$ is negative}
 \STATE Add edge $e$ to $M_i$
 \ENDIF
 \ENDFOR
\IF{$| \mathcal{F}_{\Sigma}| < K$}
 \STATE $ \mathcal{F}_{\Sigma}[c] = |M_i|$
 \STATE $\mathcal{E}_{\Sigma}[c] = M_i$
 \STATE Fetch $H_i$ by executing Algorithm~\ref{alg-Har} with inputs $\Sigma$ and $M_i$
 \STATE $\mathcal{H}_{\Sigma}[c]$=$H_i$
 \STATE $c$ = $c$ + 1
 \ELSE
 \STATE Get index $l$ such that $ \mathcal{F}_{\Sigma}[l]$ is the largest
 \IF{$ \mathcal{F}_{\Sigma}[l] < |M_i|$}
 \STATE Delete $ \mathcal{F}_{\Sigma}[l]$, $\mathcal{E}_{\Sigma}[c]$, and $\mathcal{H}_{\Sigma}[c]$
 \STATE $ \mathcal{F}_{\Sigma}[l] = |M_i|$
 \STATE $\mathcal{E}_{\Sigma}[l] = M_i$
 \STATE Fetch $H_i$ by executing Algorithm~\ref{alg-Har} with inputs $\Sigma$ and $M_i$
 \STATE $\mathcal{H}_{\Sigma}[l]$=$H_i$
 \ENDIF
\ENDIF

\ENDFOR
\STATE Return $ \mathcal{F}_{\Sigma}$, $\mathcal{E}_{\Sigma}$, and $\mathcal{H}_{\Sigma}$
\end{algorithmic}
\end{algorithm}

The ABCD Phase 1 generates the nearest balanced states, retrieves the top-$K$ states with the lowest frustration, and identifies the candidate edges causing an imbalance of each $K$ state by comparing the edge signs before and after the balancing process. $I$ is the number of iterations we run the algorithm and the upper bound on how many optimal balanced states we discover in the process. The Algorithm~\ref{alg-ABCD1} outlines the Phase 1 steps in detail. We also get the placement of each vertex in the Harary subsets of the vertices along each candidate edge of $K$ states. Essentially, we (1) discover the fundamental cycle bases for each of the $I$ spanning trees; (2) for each of the cycles in the basis, count the number of cycles that contain the odd number of negative edges; and (3) keep only the $K$, $K <<I$ balanced states out of $I$ accessed that have the smallest number of fundamental cycles with an odd number of negative edges (imbalanced cycles).
 
\subsection{ABCD Phase 2} 
\label{ssec-ABCD2}

The ABCD Phase 2 employs an innovative vertex deletion approach for all $K$ discovered balanced states to minimize the number of vertices removed along the edges that switched signs (causing imbalance) from the graph, which increases the vertex cardinality of the largest balanced sub-graph. A graph is said to be connected if there is a path between any two vertices. Removing a vertex can potentially disconnect the graph, thus significantly reducing the size of the largest sub-graph. Studying the connectivity before and after the removal of vertices gives us insights into the critical points that maintain the graph's connectivity. In graph theory, a bridge, cut-edge, or cut arc is an edge whose deletion increases the graph's number of connected components. Removing the non-bridge vertices over bridge vertices increases the chances of graph connectivity in the vertex removal process. Detecting bridges takes O($|V|+|E|$) if we use the efficient Tarjan's algorithm \cite{tarjanbridge}. This approach is too costly for $Fr(\Sigma)$ times in the edge deletion process and prohibitive for large graphs. The total complexity will be O($Fr(\Sigma)*(|V|+|E|)$). In real graphs, $|E|>|V|$, so we approximate the complexity as O($Fr(\Sigma)*(E)$), which is too expensive for large graphs. 

\begin{algorithm}[!ht]
\caption{ABCD Phase 2}
\label{alg-ABCD2} 
\begin{algorithmic}[1] 
\STATE Input $\Sigma$, $\mathcal{E}_{\Sigma}$, $K$, $\mathcal{H}_{\Sigma}$, and integer $app$ (1 for ABCDD, 2 for ABCDH, 3 for ABCDS)
\STATE Compute the adjusted status $\mathcal{O}_{\Sigma}$ (as in Algorithm~\ref{alg-adjusted}) and the degree for each vertex (degree[] array)
\STATE Fetch the sum of neighborhood degree $nei$ by executing Algorithm~\ref{alg-neighbors}
\FOR{$i$ = 0; $i$++; $i$ < $K$}
\STATE Initialize empty set $ \mathcal{V_i}$ = $ \emptyset $
 \FOR{edges $e$, $e \in \mathcal{E}_{\Sigma}[i]$}
 \IF{any of the vertices along $e$ $\in$ $ \mathcal{V}_i$}
 \STATE Skip iteration
 \ENDIF
 \IF{$app$==2}
 \IF{$e$ is positive}
 \STATE Append the vertex of index $q$ along $e$ that has a lower sum of neighborhood degrees $nei[q]$ to set $\mathcal{V}_i$ 
 \ELSE
 \STATE Append the vertex of index $w$ along $e$ where $\mathcal{H}_{\Sigma}[i][w] = 0$ to set $ \mathcal{V}_i$
 \ENDIF
 \ELSIF{$app$==3}
 \STATE Fetch the adjusted status of both vertices of index $q$ and $w$ along edge $e$
 \IF{$\mathcal{O}_{\Sigma}$[$q$] $<$ $\mathcal{O}_{\Sigma}$[$w$]}
 \STATE Append the vertex of index $q$ along $e$ to set $ \mathcal{V}_i$ 
 \ELSE
 \STATE Append the vertex of index $w$ along $e$ to set $ \mathcal{V}_i$ 
 \ENDIF
 \ELSE
 \STATE Fetch the degree of both vertices of index $q$ and $w$ along edge $e$
 \IF{degree[$q$] $<$ degree[$w$]}
 \STATE Append the vertex of index $q$ along $e$ to set $ \mathcal{V}_i$ 
 \ELSE
 \STATE Append the vertex of index $w$ along $e$ to set $ \mathcal{V}_i$ 
 \ENDIF
 \ENDIF
\ENDFOR
\STATE Create $\Sigma'_i$ as $(V \setminus \mathcal{V}_i, E \setminus \mathcal{E}_{\Sigma}[i])$
\STATE Find the largest connected component of $\Sigma'_i$ as $\Sigma''_i$
\STATE Push $\Sigma''_i$ to $\mathcal{ABCD}$
\ENDFOR
\STATE Find the largest $\Sigma''_i \in \mathcal{ABCD}, i \in [1,K]$ and return it. 
\end{algorithmic}
\end{algorithm}

Here, we propose THREE efficient approximations of connectivity that rely on the scale-free characteristic of the prominent real graphs \cite{konect, 2016Amazon2}. Those graphs have a degree distribution that follows a power law, at least asymptotically. For example, in Table~\ref{tab-KonectScale}, WikiPolitics and WikiConflict have a relatively large max degree of 20,153 and 10,715, respectively, where the median and average degrees are much lower. Recent interest in heavy-tailed degree distribution in social, biological, and economic networks shows that a few power users connect directly or indirectly to most vertices. A large number of vertices connect to a few power users \cite{scale-free} and recently analyzed real signed graphs exhibit scale-free behavior \cite{2022Survey}. 

For every set of edges we save in the $\mathcal{E}_{\Sigma}$, their cardinality is the frustration of that stable state and saved in $ \mathcal{F}_{\Sigma}$ set in Algorithm~\ref{alg-ABCD1} and it is the frustration measure for that nearest balanced state. Thus, the number of vertices erased is bound to be smaller or equal to $ \mathcal{F}_{\Sigma}[i]$ for the $i^{th}$ nearest balanced state saved as an output of the Algorithm~\ref{alg-ABCD1}. Let's consider the upper bound on the frustration and define it as in Equation~\ref{eq-upper} from the fundamental cycle balancing theory: 
\begin{equation}
\forall i, i \in [1,K] \rightarrow \mathcal{F}_{\Sigma}[i] \leq |E|-|V| + 1
\label{eq-upper}
\end{equation}
This equation stems directly from the Corollary~\ref{col-fundCycles}. The maximum number of fundamental cycles is $|E|-|V| + 1$. The worst-case scenario for finding the largest balanced sub-graph would be for the graph that requires disconnecting each of the $|E|-|V| + 1$ cycles, and none of the edges removed along these cycles share vertices. Thus, the upper bound on the number of vertices ABCD removes in Phase 2 in Algorithm~\ref{alg-ABCD2} is $|E|-|V| + 1$. Note that for the large-scale-free graphs, $|E|$ and $|V|$ are within the order of magnitude. The number of the fundamental cycles is much smaller than the number of edges, as illustrated in Table~\ref{tab:amazonData}). 

Algorithm~\ref{alg-ABCD2} summarizes Phase 2 steps for each stable state. First, we initialize an empty set $ \mathcal{V}_i$ to save the vertices from being deleted along the candidate edges (edges that flip signs). If either vertex is already in $ \mathcal{V}_i$, move on to the next edge. We add the vertex to $ \mathcal{V}_i$ based on one of the following three criteria, depending on the selection of the $app$ parameter: (1) ABCDD: Subsection~\ref{sssec-ABCDD}, $app$ = 1; (2) ABCDH: Subsection~\ref{sssec-ABCDH}, $app$ = 2; (3) ABCDS: Subsection~\ref{sssec-ABCDS}, $app$ = 3. We repeat this step $K$ times, $\forall i,$ $i$ $\in [1, K]$. The resulting outcome of the Algorithm~\ref{alg-ABCD2} is the set $V_i$ of vertices to be erased for each of the $K$ stable states where $V_i$ is going to be used to obtain the graph $\Sigma''_i$ as the largest connected component of $(V \setminus \mathcal{V}_i, E \setminus \mathcal{E}_{\Sigma}[i])$. Finally, each $K$ $\Sigma''_i$ will be stored in $\mathcal{ABCD}$, and the largest $\ Sigma'_i$ will be returned as the outcome for this task. Figure~\ref{fig-pipeline} summarizes the entire ABCD pipeline with Phase 2 containing the three handling criteria for purging vertices split by horizontal lines. 

\begin{figure*}[!ht]
 \centering
 \includegraphics[width=\linewidth]{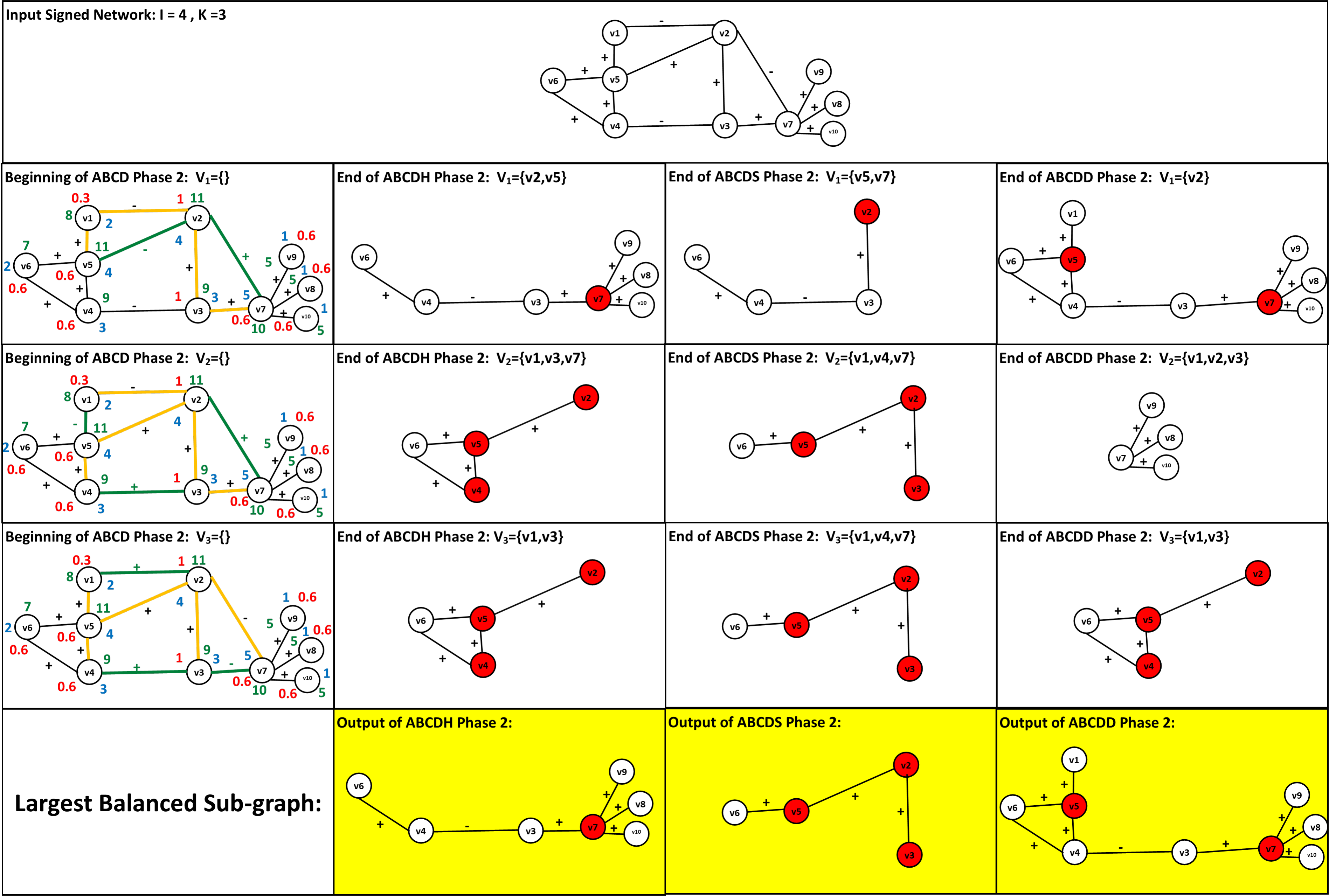}
 \caption{The ABCD algorithm applied to a sample signed graph with ten vertices and thirteen edges. For connectivity approximations, we compute the Harary bipartition in ABCD Phase 1 (Algorithm~\ref{alg-ABCD1}) and compute the sum of neighborhood (green), degree (blue), and adjusted status (red) at the beginning of Phase 2 (Algorithm~\ref{alg-ABCD1}) once for the entire graph. Orange edges are the edges of the unbalanced fundamental cycles. Green edges are the candidate edges. The result of Phase 2 for all three connectivity approximations is the sub-graph defined by black and red vertices.}
 \label{fig-ABCDH}
\end{figure*}

In the next three subsections, we will describe the three approaches in this phase that are efficient approximations for connectivity modeling. Figure~\ref{fig-ABCDH} displays three versions of ABCD Phase 2 step-by-step on a sample signed graph with seven vertices and ten edges, introduced at the top. Phase 2 starts by retrieving Harary bipartitions from Phase 1 and then calculating neighborhood sum (green), degree (blue), and adjusted status (red). Unbalanced fundamental cycle edges are shown in orange, while candidate edges are in green. At phase's end, one candidate vertex is marked for deletion and removed in the figure; the red vertices are candidate vertices that remain. The largest balanced sub-graph has the most vertices among the three produced sub-graphs.

\begin{figure}[!ht]
 \centering
 \includegraphics[width=\linewidth]{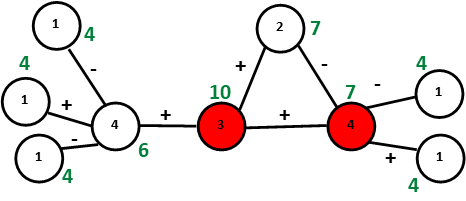}
 \vspace*{-2em}
 \caption{Degree (black, in node) vs. Sum of Neighborhood Degrees (green, next to the node) computation. The sum of neighborhood degrees labels the red vertices connected by a positive link that are candidates for deletion based on the ABCDD criteria.}
 \label{fig-neighborhood}
\end{figure}

 \begin{algorithm}[!ht]
\caption{Computation of the sum of neighborhood degree}\label{alg-neighbors}
\begin{algorithmic}[1] 
\STATE Input signed graph $\Sigma$, degree array, and $|V|$\;
\STATE Declare and initialize array neighborhood\_degree = []\;
\FOR{$q$ = 0; $q$++; $q$ < $|V|$} 
 \STATE Declare and initialize integer sum = 0\;
 \FOR{every neighbor of index $nei$ connected to vertex of index $q$ via an edge}
 \STATE sum += degree[$nei$]\;
 \ENDFOR
 \STATE neighborhood\_degree[$q$]=sum\;
\ENDFOR
\STATE Return array neighborhood\_degree\;
\end{algorithmic}
\end{algorithm}

\subsubsection{ABCDD: Algorithm for Balanced Component Discovery Degree} 
\label{sssec-ABCDD}

The algorithm for Balanced Component Discovery Degree (ABCDD) models the graph's overall connectivity as the degree. The degree of a vertex (the number of edges connected) can impact the graph's connectivity. Removing high-degree vertices removes more edges from the graph, which can introduce smaller connected components, specifically for scale-free graphs. Our goal is exactly the opposite. The approach eliminates the vertex with a smaller (neighborhood) degree out of the two. Algorithm~\ref{alg-neighbors} outlines the computation of the sum of neighborhood degrees measure. We also observe that performing the process three times where in each iteration, we set degree[] equal to neighborhood\_degree[] and then compute a new neighborhood\_degree[] using the updated degree[] generally enhances the results. Figure~\ref{fig-neighborhood} demonstrates how the sum of a neighborhood can be better in certain cases than the degree as a criterion for purging vertices. The two red vertices in the image indicate that the balancing algorithm has labeled the edge and that its sign needs to be switched for the graph to achieve a complete balancing state. The degree of the left vertex is 3, and the right vertex is 4. The neighborhood degree of the left vertex is 10 (neighbors of a neighbor), and the neighborhood degree of the right vertex is 7. We chose the vertex on the right to delete and the vertex on the left to keep (if we used degree as a criterion to delete vertices, we wouldn't have obtained the largest balanced sub-graph in this particular case). For the following experiments, we use degree and not the sum of neighborhood degrees as a handling criterion for purging vertices.

\subsubsection{ABCDH: Algorithm for Balanced Component Discovery Degree Harary} 
\label{sssec-ABCDH}

The stable states to $\Sigma$, defined in Def.~\ref{def-near}, do not have the same sets of candidate edges for balancing. The balancing can be achieved in multiple ways, as explained in detail in \cite{2021Cloud}. The proposed balancing algorithm identifies the stable states and the exact edges to remove to balance the remaining sub-graph. Algorithm~\ref{alg-Har} changes the signs of those edges and creates sets $U$ and $W$ for each stable state. Each of the sets is balanced and with positive connectivity among its vertices. If the original edge was negative, now it is positive, and both vertices are either in $U$ or $W$. Note that most real networks have ~20\% of opposing edges compared to 80\% of positive edges \cite{2021Cloud}. Thus, for the deletion criteria in the ABCDH, there is less chance for the edge to end up in one of the bipartitions. The majority case is now if the original edge was positive, and now it is negative; one vertices is in $U$, and another is in $W$. 

\begin{algorithm}
\begin{algorithmic}\caption{Harary Algorithm\label{alg-Har}}
\STATE Input $\Sigma$, set of edges that should flip their signs $|E|$.
\STATE Create zero vector $H$ of dimension $|V|$ 
\FOR{edge $e \in M$}
\STATE switch edge sign in $\Sigma$: $e^- \rightarrow e^+; e^+ \rightarrow e^-$
\ENDFOR
\STATE Cut all the negative edges to create Harary bi-partitions $U$ and $W$ so that $|U| > |W|$
\FOR{every vertex of index $q$ $\in$ $\Sigma$}
\STATE if $v \in U, H[q] = 1$
\ENDFOR
\STATE Return $H$
\end{algorithmic}
\end{algorithm}

The Algorithm for Balanced Component Discovery Degree Harary (ABCDH) approximates the connectivity by placing the vertex from the smaller set $W$ in candidate deletion set $ \mathcal{V_i}$ if the edge is negative. If the original edge is negative, the balancing algorithm will switch to positive and place both vertices in the same Harary partition. In that case, we resort to the ABCDD baseline, where the sum of neighborhood degrees determines which vertices to delete. If both have the same sum of neighborhood degrees, choose a random vertex along that edge to discard. Note that the \emph{neighborhood} degree is computed for all vertices in the signed graph \emph{once} and re-used for computation. 

\subsubsection{ABCDS: Algorithm for Balanced Component Discovery Status} 
\label{sssec-ABCDS}

\begin{algorithm}[!ht]
\caption{Adjusted Status Computation}
\label{alg-adjusted} 
\begin{algorithmic}[1] 
\STATE Input $\Sigma$, $\mathcal{H}_{\Sigma}$, and $K$
\STATE Initialize zero array $\mathcal{O}$ of size $|V|$
\FOR{$i$ = 0; $i$++; $i$ < $K$}
 \FOR{every vertex of index $x$ $\in$ $\Sigma$}
 \STATE $\mathcal{O}_{\Sigma}[x]$ += $\mathcal{H}_{\Sigma}[i][x]$
\ENDFOR
\ENDFOR
\FOR{every vertex of index $x$ $\in$ $\Sigma$}
 \STATE $\mathcal{O}_{\Sigma}[x]$ /= $K$
\ENDFOR
\STATE Return $\mathcal{O}$
\end{algorithmic}
\end{algorithm}
The Algorithm for Balanced Component Discovery Degree Status (ABCDS) approximates the connectivity by placing the vertex with the smaller adjusted status measure of the two vertices along an edge that switched sign in the candidate deletion set $ \mathcal{V_i}$. The ABCDS uses the adjusted status measure from \cite{2021Cloud} to determine which vertex to delete. The adjusted status computation takes all $K$ binary $\mathcal{H}_{\Sigma}$ vectors that Algorithm~\ref{alg-ABCD1} created and sums all $K$ $\mathcal{H}_{\Sigma}$ element-wise, and divides each element in this array by $K$ to define the adjusted status of the vertex. The logic behind using this adjusted status using $K$ balanced state's Harary bipartition is that it is more robust than using a single Harary bipartition as shown in \cite{2021Cloud}, which captures a node's importance, and it is easy to compute. Algorithm~\ref{alg-adjusted} outlines the steps for computing this adjusted status using the Harary bipartition binary vectors. If the statuses of both vertices are the same, choose a random vertex along that edge to discard.

\section{Complexity Analysis}
\label{sec-complexity}

In ABCD Phase 1, balancing the signed network for all vertices takes $O(|E|*log(|V|)*d_a)$ where $d_a$ is the average degree of a vertex as analyzed in \cite{2021Alabandi}. The time complexity to count the number of edge sign switches that occurred after is O($|E|$). ABCD Phase 2 takes O($K*(|E|*log(|V|)+|E|)$) because we have to loop over every top-$K$ balanced state. For each state, we loop over candidate edges, insert vertices to be kept in a set that takes $log(|V|)$ (assuming the set data structure in C++ is a red-black tree), and reread O($|E|$) to reread each top-$K$ balanced state without the candidate vertices and detect the largest connected component using union-find. Hence, the total time complexity of the algorithm is O($I*(|E|*log(|V|)*d_a)+K*(|E|*log(|V|)$). For each of the Phase 2 vertex deletion criteria, we add the following: 

\noindent \textbf{ABCDD:} Computing the degrees of each vertex and reading the graph take O($|E|$), assuming Harary bipartitions are not computed. So the total beginning-to-end complexity of ABCDD is O($|E|*log(|V|)*d_a$).

\noindent \textbf{ABCDH:} Obtaining the Harary bipartition takes O($|V|*|E|$) because we use the Belman-Ford algorithm to compute distances on the detected connected components. In case the number of balanced states retrieved is less than $K$, the time complexity for storing the state and Harary bipartition takes O($|E|$). On the other hand, when the number of stable states exceeds $K$, finding the balanced state with the largest edge sign switches and replacing it with a state of lower frustration takes O($|E|$) in total as well. Moreover, computing the degrees of each vertex and reading the graph take O($|E|$). Computing the sum of the neighborhood degrees of each vertex uses the same procedure as calculating the degrees, but it's repeated three times. Hence, O($|E|$) is added. So, the total beginning-to-end complexity of ABCDH is O($|E|*log(|V|)*d_a + |V|*|E|$). 

\noindent \textbf{ABCDS:} The total time complexity is similar to ABCDH as we sum all $K$ $\mathcal{H}_{\Sigma}$ element-wise, and divide each element in this array by $K$ which takes O($K*|V|$) as we have pre-computed the Harary bipartitions. However, it won't be a dominant term, so the total beginning-to-end complexity of ABCDS is the same as ABCDH, which is O($|E|*log(|V|)*d_a + |V|*|E|$).

\begin{table*}[!ht]
\setlength\tabcolsep{1pt}
\centering
\caption{Konect plus Twitter Ref. and PPI properties. Avg, Median, and Max refer to the degree.}
\label{tab-KonectScale}
\begin{tabular}{cccccccccc}
\hline
\textbf{Signed Graph} & \textbf{ \# vertices}&\textbf{\# edges}& \textbf{\# cycles} &\textbf{Density} & \textbf{\# of Triads} & \textbf{Avg} & \textbf{Median} & \textbf{Max} &\textbf{\% of $e^-$} \\
\hline
Highland & 16 & 58 & 43& 0.483 & 68 &7.25& 7.5& 10 &50 \\
CrisisInCloister & 18 & 126 & 145 & 0.82 & 479 & 14 &14 &17 & 42.06 \\
ProLeague & 16 & 120 & 105 & 1.0 & 560 & 15.0 &15 & 15&10.83 \\
DutchCollege &32 &422 &391 & 0.85 & 3,343 & 26.37 & 28& 31& 0.47 \\
Congress & 219 & 521 &303 & 0.021 & 212 & 4.71 &3 &33 & 20.34 \\
PPI & 3,058 & 11,860 & 8,803 & $<$ 0.01 & 3,837 & 3.87 &2 &55 & 32.5 \\
BitcoinAlpha &3,775 &14,120 &10,346 & $<$ 0.01 & 22,153 & 7.48 & 2& 511& 8.39 \\
BitcoinOTC & 5,875 & 21,489 & 15,615 & 0.01 & 33,493 & 7.31 &2 &795 & 13.57 \\
Chess & 7,115 & 55,779 & 48,665 &$<$ 0.01 & 108,584 & 15.67 & 7 &181 & 24.15\\
TwitterReferendum & 10,864& 251,396 & 240,533 &$<$ 0.01 & 3,120,811 & 46.28 & 12&2,784 & 5.08 \\
SlashdotZoo &79,116 &467,731 & 388,616 & $<$ 0.01 & 537,997 & 11.82 &2 &2,534 & 25.16 \\
Epinions & 119,130 & 704,267 & 585,138 & $<$ 0.01 & 4,910,009 & 11.82 & 2&3,558 & 16.82 \\
WikiElec &7,066& 100,667 & 93,602 & $<$ 0.01 & 607,279 & 28.49 &4 &1,065 & 21.94 \\
WikiConflict &113,123 &2,025,910 & 1,912,788 & $<$ 0.01 &13,852,201 & 35.81 & 4& 20,153& 62.33 \\
WikiPolitics & 137,740 & 715,334 & 577,595 & $<$ 0.01 & 2,978,021 & 10.38 &2 & 10,715& 11.98 \\
\hline
\end{tabular}
\end{table*}

\section{Proof Of Concept}
\label{sec-Exp}

All real-world benchmark graphs have one significant connected component, and the {implementation} of the algorithms is in C++. The algorithm identifies the largest connected component (LCC). It applies the ABCD algorithm to the largest connected component. The implementation treats edges without signs as positive edges in the fundamental cycle. ABCD Phase 1 (Algorithm~\ref{alg-ABCD1}) implementation builds on \cite{2021Alabandi} and has recently shown that the breadth-first search sampling of the spanning trees captures the diversity of the frustration cloud \cite{2021Cloud}. We adopt the breath-first search method for sampling spanning trees in Algorithm~\ref{alg-ABCD1}. Algorithm~\ref{alg-ABCD2} implements ABCD Phase 2, and the final set of largest connected components per stable states is returned in the $\mathcal{ABCD}$ set, and the winner is the largest among them. 

\FloatBarrier
\begin{sidewaystable*}
\setlength\tabcolsep{1pt}
\centering
\caption{Comparison of TIMBAL and ABCD $I$ = 5000 runs on Konect plus TwitterReferendum and PPI signed graphs. The numbers between the parenthesis in the ABCDH column are the sub-graph \# vertices and time for the faster version of ABCDH with different parameters. t stands for time.} 
\label{tab-konect}
\begin{tabular}{lllllcrrrrrr}\toprule
& \multicolumn{2}{c}{\bf TIMBAL 5 runs} & \multicolumn{2}{c}{\bf TIMBAL 10 runs} & \multicolumn{2}{c}{\bf ABCDH}& \multicolumn{2}{c}{\bf ABCDS}& \multicolumn{2}{c}{\bf ABCDD}\\ 
 \textbf{\bf Konect} & \multicolumn{1}{c}{\bf sub-graph} &\multicolumn{1}{c}{\bf Run} & \multicolumn{1}{c}{\bf sub-graph} &\multicolumn{1}{c}{\bf Run} & \multicolumn{1}{c}{\bf sub-graph} & \multicolumn{1}{c}{\bf Run} & \multicolumn{1}{c}{\bf sub-graph} & \multicolumn{1}{c}{\bf Run}& \multicolumn{1}{c}{\bf sub-graph} & \multicolumn{1}{c}{\bf Run} \\
 \textbf{Dataset} &\multicolumn{1}{c}{ \bf \# vertices}& \multicolumn{1}{c}{\bf t (s)} & \multicolumn{1}{c}{ \bf \# vertices}& \multicolumn{1}{c}{\bf t (s)} &\multicolumn{1}{c}{\bf \# vertices}& \multicolumn{1}{c}{\bf t (s)} &\multicolumn{1}{c}{\bf \# vertices}& \multicolumn{1}{c}{\bf t (s)} &\multicolumn{1}{c}{\bf \# vertices}& \multicolumn{1}{c}{\bf t (s)} \\ \midrule
 \emph{Highland}& 13 & 0.1 & 13 & 0.2 & 13 (13) & 0.9 (0.18) &12 & 1.07&12 &0.98\\ 
 \emph{CrisisInCloister}& 8 & 0.25& 8 & 0.5 & 8 (8) & 1.28 (0.25) & \textbf{9} & 1.8&7 &1.5 \\ 
 \emph{ProLeague} & 10 & 0.1 & 10 &0.2 & 10 (10) & 1.40 (0.28) & 10 & 1.6&7 &1.32\\
 \emph{DutchCollege}& 29 & 0.1 & 29 & 0.2 & \textbf{30} (30) & 14 (2.5) & \textbf{30} & 14.1&\textbf{30} &14.98 \\ 
 \emph{Congress}& 207 & 0.85 & 207 & 1.9 & 207 (207) & 7.79 (1.44)& 206 & 2.38&\textbf{210} & 9.03
\\ 
 \emph{PPI}& 900 & 78.45 & 900 & 156.9 & 2,073 (2,033) & 97.59 (18.8) & 2,038& 99.48& \bf 2,149&105.46\\ 
 \emph{BitcoinAlpha} & 3,014 & 5.57 & 3,081 & 11.4 & 3,146 (3,107) & 55.68 (14.79) & \textbf{3,154} &231.19&3,077 &221.64\\ 
 \emph{BitcoinOTC} & 4,250 & 10.15 & 4,349 & 20.3 & \textbf{4,910} (4,890) &92.29 (23.48) & 4,814 & 350.12 &4,880 & 361.59\\ 
 \emph{Chess} & 2,230 & 15.25 & 2,320 & 30.5 & \textbf{2,551} (2,554) & 174.67 (40.72) & 2,230 & 440.98&2,199 &443.27\\ 
 \emph{TwitterReferendum}& 9,021 & 27,6 & 9,110 & 55.2 & 9,438 (9,263) & 851.94 (231.29) & \textbf{9,453} & 4397.77&8,622 &3479.59\\ 
 \emph{SlashdotZoo} & 39,905& 259.4 & 40,123 & 518.8 &43,544 (43,219) & 1870.20 (381.24) & \textbf{45,250} & 4530.95& 41,290&4426.75 \\
\emph{Epinions} & 73,433 & 774.9 & 74,106 & 1549.8 & 74,843 (74,522) & 3272.87 (632.39) & \textbf{78,126} &2406& 74,555& 2385.04\\
 \emph{WikiElec}& \textbf{3,758} & 18.95 & \textbf{3,856} & 37.9 & 3,506 (3,506) & 3033.08 (73.43) & 3,528 & 831.78& 3,142& 805.73\\ 
 \emph{WikiConflict} & 56,768 & 539.25 & 56,768 & 1078.5 & 54,476 (53,549) & 8332.22 (1979.049) & 57,748 & 8317.99& \bf 60,606&8344.92 \\ 
 \emph{WikiPolitics} & 67,009 & 602.65 & 69,050 & 1205.3 & 63,584 (63,584) &3478.08 (672.077) & 65,730 &2790.39& \bf 69,520 & 2779.16\\ \bottomrule
\end{tabular}
\end{sidewaystable*}
\FloatBarrier

\begin{figure*}[!ht]
 \centering
 \includegraphics[width=0.54\linewidth]{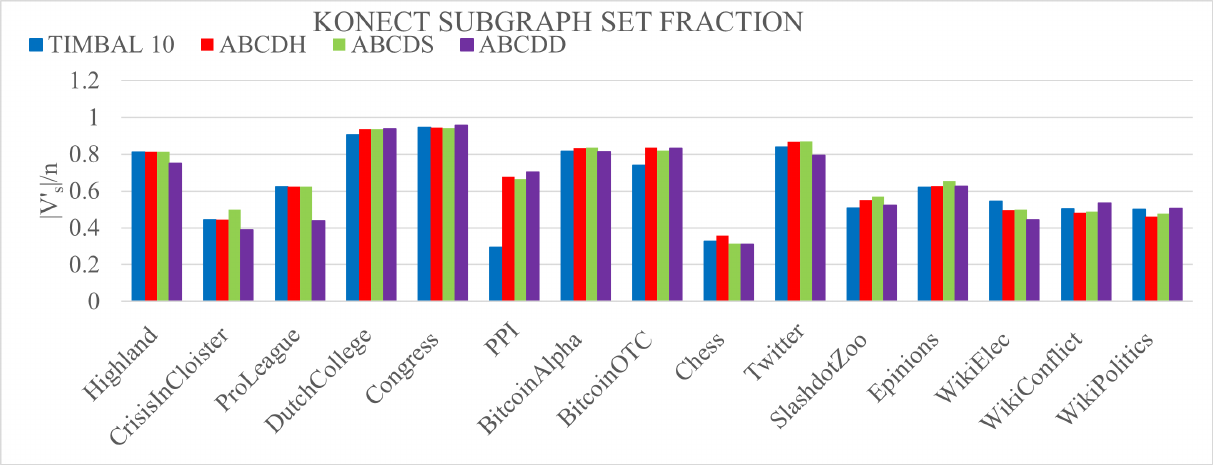}
 \includegraphics[width=0.425\linewidth]{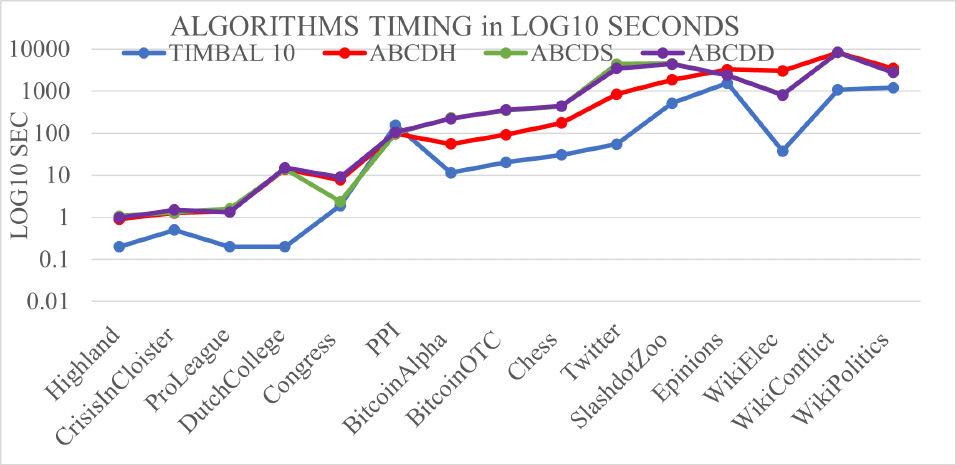}
 \caption{ABCD and TIMBAL performance comparison for Konect benchmark in terms of subset graph fractions (left) and algorithmic timing (right).}
 \label{fig-konect_abcd}
\end{figure*}

\textbf{Baseline:} The TIMBAL approach has reached the highest cardinality of the sub-graphs in various datasets and is a de-facto state-of-the-art \cite{TIMBAL}. The input parameters of TIMBAL \cite{TIMBAL} are set as follows for all subsample\_flag=False, samples=4 based on the paper implementation. The parameter max\_removals=1 is set for small graphs (under 1000 vertices) and to max\_removals=100 for the rest of the signed networks. e set avg\_size=20 for datasets of several vertices less than 80,000, and subsample\_flag=True, samples = 1000, avg\_size = 200 max\_removals=100 for datasets with the number of vertices greater than 80,000. TIMBAL is a non-deterministic algorithm, and we run it 5 and 10 times for Konect data to get the maximum vertex cardinality.

\textbf{Setup:} The operating system used for the experiments is Linux Ubuntu 20.04.3, running on the 11th Gen Intel (R) Core (TM) i9-11900 K @ 3.50GHz with 16 physical cores. It has one socket, two threads per core, and eight cores per socket. The architecture is X86\_x64. Its driver version is 495.29.05, and the CUDA version is 11.5. The cache configuration is L1d: 384 KiB, L1i: 256 KiB, L2: 4 MiB, L3: 16 MiB. The CPU op is 32-bit and 64-bit. 

\textbf{Comparison:} We compare three implementations of the ABCD algorithm (ABCDH, ABCDS, ABCDD) to TIMBAL for 14 Konect (plus Twitter Ref. and PPI signed graphs) and 17 Amazon datasets in terms of runtime in seconds and the size of the produced sub-graph and verify the balanced state of the discovered sub-graph for both methods. 

\textbf{ABCD} algorithm parameters: $I=5000$ for all, $K=4000$ for $n < 100,000$, $K=100$ for $100,000 < n < 300,000$, and $K=20$ for $300,000 < n$ vertices. \textbf{ABCDH\_Fast} is a faster version of \textbf{ABCDH} and the parameters are: $I=1000$ for all, $K=700$ for $n<100,000$, $K=100$ for $100,000 < n < 300,000$, and $K=20$ for $300,000 < n$ vertices. For this faster version, we can also study the effect of decreasing the number of iterations on the overall speed and performance. 

\begin{table}[!ht]
\centering
 \caption{Comparison between TIMBAL and ABCD on Amazon ratings and reviews \cite{2016Amazon2} mapped to signed graphs. The number of vertices, edges, and cycles reflect the number in the largest connected component of each dataset.}
 \label{tab:amazonData}
 \begin{tabular}{cccccc}\toprule
 \bf AMAZON & \multicolumn{1}{c} {\bf Input graph} & \multicolumn{3}{c} {\bf Largest Connected Component}\\
 \bf Ratings & \bf \# ratings& \bf \# vertices& \bf \# edges & \bf \# cycles \\ \midrule
 Books & 22,507,155& 9,973,735 & 22,268,630 & 12,294,896 \\ 
 Electronics & 7,824,482& 4,523,296 & 7,734,582 & 3,211,287 \\ 
 Jewelry & 5,748,920& 3,796,967 & 5,484,633 & 1,687,667 \\ 
 TV & 4,607,047&2,236,744 & 4,573,784 & 2,337,041 \\ 
 Vinyl & 3,749,004& 1,959,693 & 3,684,143 & 1,724,451 \\ 
 Outdoors & 3,268,695 & 2,147,848 & 3,075,419 & 927,572 \\ 
 AndrApp & 2,638,172 & 1,373,018 & 2,631,009 & 1,257,992 \\ 
 Games & 2,252,771 & 1,489,764 & 2,142,593 & 652,830\\
 Automoto & 1,373,768 & 950,831 & 1,239,450 & 288,620\\ 
 Garden & 993,490& 735,815 & 939,679 & 203,865 \\ 
 Baby & 915,446& 559,040 & 892,231 & 333,192 \\ 
 Music & 836,006 & 525,522 & 702,584 & 177,063 \\ 
 Video & 583,993 & 433,702 & 572,834 & 139,133 \\ 
 Instruments & 500,176& 355,507 & 457,140 & 101,634 \\ 
 \midrule
 \bf Reviews & \bf \# reviews & \bf \# vertices& \bf \# edges & \bf \# cycles \\ \midrule
 Core Music & 64,706& 9,109 & 64,706 & 55,598 \\ 
 Core Video & 37,126& 6,815 & 37,126 & 30,312 \\ 
 Core Instrum & 10,621 & 2,329 & 10,261 & 7,933 \\ \bottomrule
 \end{tabular}
\end{table}

\begin{table*}[!ht]
\setlength\tabcolsep{1pt}
\centering
 \caption{Comparison between TIMBAL and ABCD on Amazon ratings and reviews \cite{2016Amazon2} mapped to signed graphs. The time for ABCD is for $I$ = 5000 iterations, and the time for TIMBAL is for all runs. N/A - TIMBAL DOES NOT COMPLETE WITHIN 48 hours. t stands for time.}
\begin{tabular}{cccccccccc}\toprule
 {\bf AMAZON} & \multicolumn{2}{c}{\bf TIMBAL 1 run} &\multicolumn{2}{c}{\bf ABCDH} & \multicolumn{2}{c}{\bf ABCDS}& \multicolumn{2}{c}{\bf ABCDD}\\
\textbf{Ratings} & {\bf\# vertices} &{\bf t (hr)}& {\bf\# vertices} &{\bf t (hr)}& {\bf\# vertices} &{\bf time (hr)}& {\bf\# vertices} &{\bf t (hr)} \\ \midrule
 Books & N/A & N/A & 7,085,285&32.5&6,265,058& 32.97 &\bf 7,458,256 &32.85\\
Electronics & N/A & N/A& 3,104,399 &10.5& 2,031,543& 10.62 &\bf 3,689,985 & 10.67\\ 
 Jewelry & 530,363 &13.1& 2,769,431 & 6&2,237,260 & 5.96 & \bf 2,949,384&6.05\\ 
 TV & 891,106 &3.16& 1,579,760& 4.76 &1,299,795 & 4.84& \bf 1,795,706&4.84\\
Vinyl & 612,700&3.2 & 1,452,496 &3.61 &1,358,541& 3.70&\bf 1,474,459 &3.68\\ 
 Outdoors & 683,846 &3.53 & 1,640,544& 3.14&1,400,498& 3.16& \bf 1,823,824&3.17\\
AndrApp & 437,740 &1.4 & 977,536& 3.4&636,566 & 3.42&\bf 1,133,649 &3.45\\
 Games & 565,301 &1.74 & 1,150,782&2.12&1,042,898 & 2.16& \bf 1,261,748& 2.17 \\ 
 Automoto & 140,711 & 3.61 &744,474 & 1.15& 685,805& 1.17&\bf 801,708 &1.18\\ \midrule
 \textbf{Ratings} & {\bf \# vertices } & {\bf t (min)}& {\bf \# vertices } & {\bf t (min)}& {\bf \# vertices } & {\bf t (min)} & {\bf \# vertices } & {\bf t (min)} \\ \midrule
Baby & 229,545 & 60 & 397,940&50&300,996 & 50.34&\bf 468,446 &50.67\\
 Music & 351,124 & 53.7& 451,320&36.7&428,561 & 37.53& \bf 471,928&37.15\\
Video & 121,694 &71.3& 360,665 &36.2& 318,484& 2179.59& \bf 401,236&36.7\\
Instruments & 97,486 & 30 &285,233&24.4&273,250 & 24.55&\bf 313,811 &24.89\\ \midrule
 \textbf{Reviews} & {\bf \# vertices} &{\bf t (s)}& {\bf \# vertices} &{\bf time (s)}& {\bf \# vertices} &{\bf t (s)} & {\bf \# vertices} &{\bf t (s)} \\ \midrule
Core Music & 4,193 &30.3 & \textbf{5,143}&200.4&4,963 & 548.64&3,595 &527.18\\ 
 Core Video & 3,419 & 23.7& \textbf{3,934}&128.3&3,740 & 322.12&2,552 &318.3\\
 Core Instrum & \textbf{1,725} &19.1 & 1,559 &36.9&1,535& 110.27&1,272 &98.31\\ \bottomrule 
\end{tabular}
\label{tab-amazon}
 \end{table*}

\subsection{TIMBAL vs. ABCD for the Konect Benchmark}
\label{ssec-KonectResults}

Konect signed graphs are from \cite{konect}, and their characteristics are described in Table~\ref{tab-KonectScale}. \emph{Highland} is the signed social network of tribes of the Gahuku\-Gama alliance structure of the Eastern Central Highlands of New Guinea, from Kenneth Read. \emph{CrisisInCloister} is a directed network that contains ratings between monks related to a crisis in an abbey (or monastery) in New England (USA), which led to the departure of several of the monks. \emph{ProLeague} are results of football games in Belgium from the Pro League in 2016/2017 in the form of a directed signed graph. Vertices are teams; each directed edge from A to B denotes that team A played at home against team B. The edge weights are the goal difference, and thus favorable if the home team wins, negative if the away team wins, and zero for a draw. \emph{DutchCollege} is a directed network that contains friendship ratings between 32 first-year university students (vertices) who mostly did not know each other before starting university. Students rate each other at seven different time points. An edge between two students shows how the reviewer rated the target, and the edge weights show how good their friendship is in the eye of the reviewer. The weight ranges from -1 for the risk of conflict to +3 for best friend.
\emph{Congress} is a signed network where vertices are politicians speaking in the United States Congress, and a directed edge denotes that a speaker mentions another speaker. In the \emph{Chess} network, each vertex is a chess player, and a directed edge represents a game with the white player having an outgoing edge and the black player having an ingoing edge. The weight of the edge represents the outcome. \emph{BitcoinAlpha} is a user-user trust/distrust network from the Bitcoin Alpha platform for Bitcoin trading. \emph{BitcoinOTC} is a user-user trust/distrust network from the Bitcoin OTC platform for Bitcoin trading. \emph{TwitterReferendum} captures data from Twitter concerning the 2016 Italian Referendum. Different stances between users signify a negative tie, while the same stances indicate a positive link \cite{Lai2018}. 

\emph{WikiElec} is the network of users from the English Wikipedia that voted for and against each other in admin elections. \emph{SlashdotZoo} is the reply network of the technology website Slashdot. Vertices are users, and edges are replies. The edges of \emph{WikiConflict} represent positive and negative conflicts between users of the English Wikipedia. \emph{WikiPolitics} is an undirected signed network that contains interactions between the users of the English Wikipedia that have edited pages about politics. Each interaction, such as text editing and votes, is valued positively or negatively. \emph{Epinions} is the trust and distrust network of Epinions, an online product rating site. It incorporates individual users connected by directed trust and distrust links. \emph{PPI} models the protein-protein interaction network \cite{SSSNET2021}.

The first benchmark consists of 14 signed graphs from the Konect repository \cite{konect} (plus TwitterRef. and PPI) used in TIMBAL benchmark evaluations. Figure~\ref{fig-konect_abcd} and Table~\ref{tab-konect} summarizes baseline TIMBAL and proposed ABCD performance. The ABCD matches TIMBAL performance in two networks (Highland and Proleague). ABCD algorithm finds a more significant subset for 13 Konect datasets than TIMBAL. On the contrary, TIMBAL performs better on only one Konect signed graph (WikiElec). TIMBAL is faster than ABCD on smaller networks. For the most extensive graph in the collection, Epinions, ABCD takes double the time to recover the largest balanced sub-graph. 

We recorded the maximum number of vertices obtained after 5 and 10 runs for TIMBAL, and only for one dataset did the repeated runs discover a more significant subset. Table~\ref{tab-konect} also summarizes the results of the ABCDH\_Fast performance in the parenthesis in the ABCDH column. For this benchmark, ABCD algorithms outperform TIMBAL with comparable runtimes. 

\begin{figure*}[!t]
 \centering
 \includegraphics[width=0.5\linewidth]{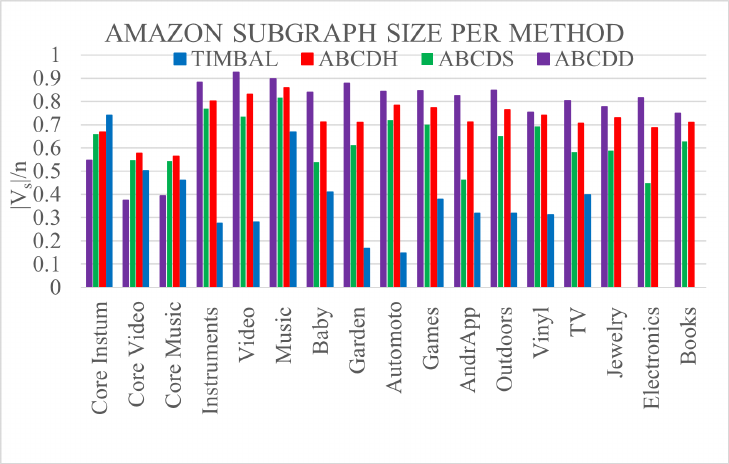}
 \includegraphics[width=0.485\linewidth]{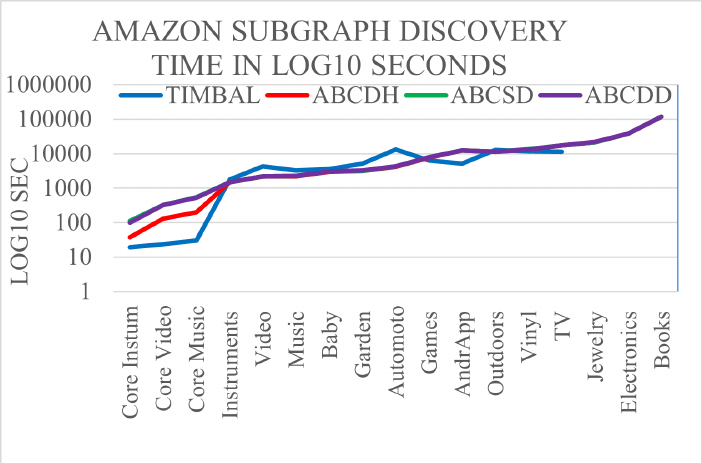}
 \caption{ABCD and TIMBAL performance and running time comparison for Amazon data.}
 \label{fig-amazonResults}
\end{figure*}

\subsection{TIMBAL vs. ABCD for the Amazon Benchmark}
\label{ssec-AmazonResults}
Amazon benchmark consists of 17 signed graphs derived from the Amazon rating and review files \cite{2016Amazon2}. The dataset contains product reviews and metadata from Amazon, spanning May 1996 to July 2014. Rating score is mapped into an edge between the user and the product as follows $(5,4) \rightarrow e^+$, $3 \rightarrow e$ (no sign), and $(2, 1) \rightarrow e^-$ \cite{2016Amazon2}. 

Table~\ref{tab:amazonData} summarizes Amazon data used and the characteristics of the largest connected component of the graph. Figure~\ref{fig-amazonResults} (left) and Table~\ref{tab-amazon} illustrate the sub-graph size TIMBAL recovers (blue box) and the sub-graph ABCD algorithm recovers (red box). Amazon data is extensive. The ABCD algorithm performs much better for millions of vertices than TIMBAL, especially the ABCDD version. One iteration of TIMBAL (blue line) takes as long as the entire ABCD algorithm (red line) for larger graphs. In this experiment, the ABCD algorithm has a superior runtime and performance regarding the graph size it discovers, as illustrated in Figure~\ref{fig-amazonResults} (right). TIMBAL's performance degrades with the graph size, and the discovered sub-graphs are much smaller than what ABCD finds, as described in Figure \ref{fig-amazonResults} (left).

\begin{table}[!ht]
\setlength\tabcolsep{5pt}
\caption{Ablation Study of ABCDH on several signed graphs with varying $K$ with $I = 1000$.}
\label{tab:paramstudy}
\centering 
\begin{tabular}{lrrrrrr} \toprule
Data& \multicolumn{2}{c}{\bf Epinions} & \multicolumn{2}{c}{\bf SlashdotZoo} & \multicolumn{2}{c}{\bf TwitterRef.}\\ 
$K$& \# vertices& time(s) & \# vertices & time(s) & \# vertices & time(s)\\ \midrule
1 & 72,417 & 448.21&41,919&286.24& 8,965 & 147.3\\ 
2 & 72,895 & 449.65 & 43,171&287.61 & 9,255 & 147.76\\ 
3 & 73,664 & 454.02 & 43,189&288.09 &9,255 & 148.25 \\
4 & 73,664 & 453.52 & 43,189&289.34 & 9,255 & 149.16 \\ 
5 & 73,664 & 454.73 & 43,189&289.95 & 9,255 & 150.53\\ 
10 & 74,522 & 467.15 &43,189&294.67& 9,255 & 155.18\\ 
20 & 74,522 & 487.90&43,189&304.84&9,263 & 163.93 \\ 
30 & 74,522 & 507.6 &43,189&315.96& 9,263 & 172.45 \\
40 & 74,522 &523.80 &43,189&324.53& 9,263 & 181.23 \\ 
50 & 74,522 & 544.65 &43,189&335.62& 9,263 & 189.6\\ \bottomrule
\end{tabular}
\end{table}

 \begin{table}[!ht]
\setlength\tabcolsep{1pt}
\caption{Ablation Study of ABCDH on several signed graphs with varying $I$ with $K = 5$.}
\label{tab:paramstudy1}
\centering \footnotesize
\begin{tabular}{lrrrrrr}\toprule
& \multicolumn{2}{c}{\bf Epinions}& \multicolumn{2}{c}{\bf SlashdotZoo}& \multicolumn{2}{c}{\bf Twitter Ref.}\\ 
$I \downarrow$ & \# vertices & time(s) & \# vertices & time(s) & \# vertices & time(s) \\ \midrule
10 & 74,209 & 18.45 & 43,219 & 10.36 & 9,161 & 7.38 \\ 
1000 & 73,664 & 455.1 & 43,189 & 292.57 & 9,255 & 160.59 \\ 
2000 & 74,053 & 887.83 & 43,338 & 568.67 & 9,200 & 293.51\\
4000 & 73,717& 1750.01 & 43,885 & 1130.36 &9,243& 593.98\\
5000 & 72,949 & 22190.82 & 43,154 & 1411.77 & 9,193 & 751.85 \\\bottomrule
\end{tabular}
\end{table}

\subsection{ABCD Ablation Study}
\label{ssec-param}

First, we select the ABCDH version of ABCD and study the effect of the two parameters $I$ and $K$ on the balanced sub-graph size. First, we set $I$ to 1000 and vary $K$ from 1 to 5 and then $K \in \{10,20,30,40,50\}$ on Epinions, SlashdotZoo, and TwitterReferendum, and Table~\ref{tab:paramstudy} summarizes the results. The ABCD found a larger balanced sub-graph of vertex cardinality 74,522 when increasing $K$ to 10 for Epinions. When we vary the number of iterations $I$ with $K$ = 5, the results in Table~\ref{tab:paramstudy1} show a reduction in the size of the largest balanced sub-graph found with a greater value of $I$. Thus, the vertex cardinality of the discovered sub-graph increases when the $K$ is larger. For $K=100$, Table~\ref{tab:paramstudy2} summarizes the improvements as the size of the sub-graph increases for comparable Table~\ref{tab:paramstudy1} performance. The execution timing also increases, and we use the size of the frustration cloud $K$ as a balancing barometer between the size of the sub-graph found and execution time. 

Next, as the number of iterations $I$ for the larger $K=100$ increases, the size of the discovered balanced sub-graph also increases for the ABCD method. The execution time also increases. However, for more iterations, the algorithm generates more. We can also observe that the more stable states we generate (the greater the value of $I$), the greater we must increase $K$ to capture the sub-graph with the largest vertex cardinality. Therefore, 

 \begin{table}[!ht]
\setlength\tabcolsep{1pt}
\caption{Ablation Study of ABCDH on several signed graphs with varying $I$ with $K = 100$.}
\label{tab:paramstudy2}
\centering \footnotesize
\begin{tabular}{lrrrrrr}\toprule
& \multicolumn{2}{c}{\bf Epinions}& \multicolumn{2}{c}{\bf SlashdotZoo}& \multicolumn{2}{c}{\bf Twitter Ref.}\\ 
$I \downarrow$ & \# vertices & time(s) & \# vertices & time(s) & \# vertices & time(s) \\ \midrule
1000 &74,522 & 634.55& 43,219 & 384.42 & 9,263 & 232.84 \\ 
2000 &74,866 & 1075.73& 43,338& 665.4& 9,276 & 374.54 \\
3000 &74,365 & 1500.92& 43,683 & 952.35 & 9,262& 521.62 \\
4000 &74,867 &1943.77 & 43,885 & 1228.63 & 9,263& 673.2 \\
5000 &74,843 &2377.75 &43,544 & 1515.50 & 9,228&830.33 \\\bottomrule
\end{tabular}
\end{table}

\section{Conclusion} 
\label{sec-Done}
Finding maximum balanced sub-graphs is a fundamental problem in graph theory with significant practical applications. While the situation is computationally challenging, the existing approximation algorithms have made considerable progress in solving it efficiently for many signed networks and propose a novel scalable algorithm for balance component discovery (ABCD). We capture the information on the unbalanced fundamental cycles and the Harary bipartition labeling for the top unique total cycle bases with the lowest number of unstable cycles. A balanced state with the lowest frustration index for a specific signed network does not necessarily yield a maximum balanced sub-graph. A balanced state with a high frustration index skyrockets the number of vertices discarded due to the increase in the number of candidate vertices and edges to be processed. We introduce a novel set of conditions (neighborhood degree, bi-cut) to remove the vertices from the graph. The output of the ABCD algorithm is guaranteed to be balanced. ABCD eliminates the unbalanced cycle bases by removing the edges. Thus, the cycle turns into an open path. The resulting sub-graph has the largest size regarding the number of vertices; it is balanced as it has no unbalanced cycles, and it is a sub-graph as the algorithm removes the \emph{vertices}. ABCD recovers significantly balanced sub-graphs over two times larger than state-of-the-art. 

Future work includes the OpenMP and GPU code accelerations as the GPU implementation of the baseline takes less than 15 minutes to find 1000 fundamental cycle bases for 10M vertices and 22M edges \cite{2021Alabandi}. Since the runtime is roughly proportional to the input size, the ABCD parallel implementation can balance ten times larger inputs in a few seconds per sample, making it tractable to analyze graphs with 100s of millions of vertices and edges.


\begin{thebibliography}{}

\bibitem[Abelson and Rosenberg, 1958]{1958Abelson}
Abelson, R.~P. and Rosenberg, M.~J. (1958).
\newblock Symbolic psycho-logic: A model of attitudinal cognition.
\newblock {\em Behavioral Science}, 3(1):1--13.

\bibitem[Alabandi et~al., 2021]{2021Alabandi}
Alabandi, G., Te\v{s}i\'{c}, J., Rusnak, L., and Burtscher, M. (2021).
\newblock Discovering and balancing fundamental cycles in large signed graphs.
\newblock In {\em Proceedings of the International Conference for High
  Performance Computing, Networking, Storage and Analysis}, SC '21, New York,
  NY, USA. Association for Computing Machinery.

\bibitem[Amelkin and Singh, 2019]{amelkin2019fighting}
Amelkin, V. and Singh, A.~K. (2019).
\newblock Fighting opinion control in social networks via link recommendation.
\newblock In {\em Proceedings of the 25th ACM SIGKDD Intl. Conf. Knowledge
  Discovery \& Data Mining}, pages 677--685.

\bibitem[Anna~D. and Aaron, 2019]{scale-free}
Anna~D., B. and Aaron, C. (2019).
\newblock Scale-free networks are rare.
\newblock {\em Nature Communications}, 10(1):1 -- 10.

\bibitem[Berger et~al., 2004]{Berger2004}
Berger, F., Gritzmann, P., and de~Vries, S. (2004).
\newblock Minimum cycle bases for network graphs.
\newblock {\em Algorithmica}, 40(1):51--62.

\bibitem[Bonchi et~al., 2019]{bonchi2019discovering}
Bonchi, F., Galimberti, E., Gionis, A., Ordozgoiti, B., and Ruffo, G. (2019).
\newblock Discovering polarized communities in signed networks.

\bibitem[Boulton, 2016]{BOULTON20161}
Boulton, L. (2016).
\newblock Spectral pollution and eigenvalue bounds.
\newblock {\em Applied Numerical Mathematics}, 99:1--23.

\bibitem[Cartwright and Harary, 1956]{Har2}
Cartwright, D. and Harary, F. (1956).
\newblock Structural balance: a generalization of {H}eider's theory.
\newblock {\em Psychological Rev.}, 63:277--293.

\bibitem[Chen et~al., 2023]{Chen2023}
Chen, C., Wu, Y., Sun, R., and Wang, X. (2023).
\newblock Maximum signed $\theta$-clique identification in large signed graphs.
\newblock {\em IEEE Transactions on Knowledge and Data Engineering},
  35(2):1791--1802.

\bibitem[Crowston et~al., 2013]{Crowstown2013}
Crowston, R., Gutin, G., Jones, M., and Muciaccia, G. (2013).
\newblock Maximum balanced subgraph problem parameterized above lower bound.
\newblock {\em Theoretical Computer Science}, 513:53 -- 64.

\bibitem[Deng et~al., 2007]{Deng_2007}
Deng, K., Zhao, H., and Li, D. (2007).
\newblock Effect of node deleting on network structure.
\newblock {\em Physica A: Statistical Mechanics and its Applications},
  379(2):714--726.

\bibitem[Derr et~al., 2020]{derr2020link}
Derr, T., Wang, Z., Dacon, J., and Tang, J. (2020).
\newblock Link and interaction polarity predictions in signed networks.
\newblock {\em Social Network Analysis and Mining}, 10(1):1--14.

\bibitem[Facchetti et~al., 2011]{frustcite}
Facchetti, G., Iacono, G., and Altafini, C. (2011).
\newblock Computing global structural balance in large-scale signed social
  networks.
\newblock {\em Proceedings of the National Academy of Sciences},
  108(52):20953--20958.

\bibitem[Figueiredo and Frota, 2014]{Figueiredo2014}
Figueiredo, R. and Frota, Y. (2014).
\newblock The maximum balanced subgraph of a signed graph: Applications and
  solution approaches.
\newblock {\em European Journal of Operational Research}, 236(2):473--487.

\bibitem[Figueiredo et~al., 2011]{Figueiredo2011}
Figueiredo, R.~M., Labbé, M., and de~Souza, C.~C. (2011).
\newblock An exact approach to the problem of extracting an embedded network
  matrix.
\newblock {\em Computers and Operations Research}, 38(11):1483 -- 1492.

\bibitem[Garimella et~al., 2021]{garimella2021political}
Garimella, K., Smith, T., Weiss, R., and West, R. (2021).
\newblock Political polarization in online news consumption.
\newblock In {\em Proceedings of the International AAAI Conference on Web and
  Social Media}, volume~15, pages 152--162.

\bibitem[G\"{u}lpinar et~al., 2004]{GULPINAR2004359}
G\"{u}lpinar, N., Gutin, G., Mitra, G., and Zverovitch, A. (2004).
\newblock Extracting pure network submatrices in linear programs using signed
  graphs.
\newblock {\em Discrete Applied Mathematics}, 137(3):359--372.

\bibitem[Harary and Cartwright, 1968]{Harary1968}
Harary, F. and Cartwright, D. (1968).
\newblock On the coloring of signed graphs.
\newblock {\em Elemente der Mathematik}, 23:85--89.

\bibitem[Harary et~al., 2002]{Harary2002}
Harary, F., Lim, M.-H., and Wunsch, D.~C. (2002).
\newblock Signed graphs for portfolio analysis in risk management.
\newblock {\em IMA Journal of Management Mathematics}, 13(3):201--210.

\bibitem[He and McAuley, 2016]{2016Amazon2}
He, R. and McAuley, J. (2016).
\newblock Ups and downs: Modeling the visual evolution of fashion trends with
  one-class collaborative filtering.
\newblock In {\em Proceedings of the 25th Intl. Conf. World Wide Web}, WWW '16,
  pages 507--517. ACM.

\bibitem[He et~al., 2021]{SSSNET2021}
He, Y., Reinert, G., Wang, S., and Cucuringu, M. (2021).
\newblock {SSSNET:} semi-supervised signed network clustering.
\newblock In {\em Proceedings of the 2022 SIAM Intl. Conf. Data Mining (SDM)},
  pages 244--252.

\bibitem[Heider, 1946]{Heider}
Heider, F. (1946).
\newblock Attitudes and cognitive organization.
\newblock {\em J.\ Psychology}, 21:107--112.

\bibitem[Interian et~al., 2022]{interian2022network}
Interian, R., Marzo, R.~G., Mendoza, I., and Ribeiro, C.~C. (2022).
\newblock Network polarization, filter bubbles, and echo chambers: An annotated
  review of measures, models, and case studies.
\newblock {\em arXiv preprint arXiv:2207.13799}.

\bibitem[Kleinberg et~al., 2008]{conn1}
Kleinberg, J., Sandler, M., and Slivkins, A. (2008).
\newblock Network failure detection and graph connectivity.
\newblock {\em SIAM Journal on Computing}, 38(4):1330--1346.

\bibitem[Kunegis, 2013]{konect}
Kunegis, J. (2013).
\newblock {KONECT} -- {The} {Koblenz} {Network} {Collection}.
\newblock In {\em Proceedings of the 22nd Intl. Conf. World Wide Web}, WWW '13,
  pages 1343--1350. ACM.

\bibitem[Lai et~al., 2018]{Lai2018}
Lai, M., Patti, V., Ruffo, G., and Rosso, P. (2018).
\newblock Stance evolution and Twitter interactions in an Italian political
  debate.
\newblock In Silberztein, M., Atigui, F., Kornyshova, E., M{\'e}tais, E., and
  Meziane, F., editors, {\em Natural Language Processing and Information
  Systems}, pages 15--27, Cham. Springer International Publishing.

\bibitem[Lalou et~al., 2018]{LALOU201892}
Lalou, M., Tahraoui, M.~A., and Kheddouci, H. (2018).
\newblock The critical node detection problem in networks: A survey.
\newblock {\em Computer Science Review}, 28:92--117.

\bibitem[Leskovec et~al., 2010]{signedsocial}
Leskovec, J., Huttenlocher, D., and Kleinberg, J. (2010).
\newblock Signed networks in social media.
\newblock In {\em Proceedings of the SIGCHI Conference on Human Factors in
  Computing Systems}, CHI '10, page 1361–1370, New York, NY, USA. Association
  for Computing Machinery.

\bibitem[Liu et~al., 2022]{Liu2022}
Liu, C., Dai, Y., Yu, K., and Zhang, Z. (2022).
\newblock Enhancing cancer driver gene prediction by protein-protein
  interaction network.
\newblock {\em IEEE/ACM Transactions on Computational Biology and
  Bioinformatics, Computational Biology and Bioinformatics, IEEE/ACM
  Transactions on IEEE/ACM Trans. Comput. Biol. and Bioinf}, 19(4):2231 --
  2240.

\bibitem[Macon et~al., 2012]{Macon2012}
Macon, K.~T., Mucha, P.~J., and Porter, M.~A. (2012).
\newblock Community structure in the United Nations General Assembly.
\newblock {\em Physica A: Statistical Mechanics and its Applications},
  391(1):343--361.

\bibitem[Ordozgoiti et~al., 2020]{TIMBAL}
Ordozgoiti, B., Matakos, A., and Gionis, A. (2020).
\newblock Finding large balanced subgraphs in signed networks.
\newblock In {\em Proceedings of The Web Conference 2020}, WWW '20, pages
  1378--1388, New York, NY, USA. Association for Computing Machinery.

\bibitem[Poljak and Turz\'{i}k, 1986]{POLJAK1986}
Poljak, S. and Turz\'{i}k, D. (1986).
\newblock A polynomial-time heuristic for certain subgraph optimization
  problems with guaranteed worst-case bound.
\newblock {\em Discrete Mathematics}, 58(1):99--104.

\bibitem[Rusnak and Te\v{s}i\'{c}, 2021]{2021Cloud}
Rusnak, L. and Te\v{s}i\'{c}, J. (2021).
\newblock Characterizing attitudinal network graphs through frustration cloud.
\newblock {\em Data Mining and Knowledge Discovery}, 6.

\bibitem[Sharma et~al., 2021]{2021Sharma}
Sharma, K., Gillani, I.~A., Medya, S., Ranu, S., and Bagchi, A. (2021).
\newblock {\em Balance Maximization in Signed Networks via Edge Deletions},
  pages 752--760.
\newblock Association for Computing Machinery, New York, NY, USA.

\bibitem[Tarjan, 1974]{tarjanbridge}
Tarjan, R. (1974).
\newblock A note on finding the bridges of a graph.
\newblock {\em Information Processing Letters}, 2(6):160--161.

\bibitem[Tomasso et~al., 2022]{2022Survey}
Tomasso, M., Rusnak, L., and Te\v{s}i\'{c}, J. (2022).
\newblock Advances in scaling community discovery methods for signed graph
  networks.
\newblock {\em Journal of Complex Networks}, 10(3).

\bibitem[Wu et~al., 2022]{2022wusurvey}
Wu, Y., Meng, D., and Wu, Z.-G. (2022).
\newblock Disagreement and antagonism in signed networks: A survey.
\newblock {\em IEEE/CAA Journal of Automatica Sinica, Automatica Sinica,
  IEEE/CAA Journal of, IEEE/CAA J. Autom. Sinica}, 9(7):1166 -- 1187.

\bibitem[Zaslavsky, 2012]{2012Zaslavsky}
Zaslavsky, T. (2012).
\newblock A mathematical bibliography of signed and gain graphs and allied
  areas.
\newblock {\em ELECTRONIC JOURNAL OF COMBINATORICS}.
\end{thebibliography}
\end{document}